\documentclass[prl,twocolumn,fleqn,showpacs,showkeys]{revtex4}
\usepackage{graphicx}
\usepackage{amssymb}
\usepackage{amsmath}
\usepackage{bm}
\begin{document}
\setcounter{page}{1}
\title[]{Photoinduced Perturbations of the Magnetic Superexchange in Core-Shell Prussian Blue Analogues}
\author{Elisabeth S. \surname{Knowles}}
\affiliation{Department of Physics and NHMFL, University of Florida, Gainesville, FL 32611-8440, USA}
\author{Matthieu F. \surname{Dumont}}
\affiliation{Department of Physics and NHMFL, University of Florida, Gainesville, FL 32611-8440, USA}
\author{Marcus K. \surname{Peprah}}
\affiliation{Department of Physics and NHMFL, University of Florida, Gainesville, FL 32611-8440, USA}
\author{Carissa H. \surname{Li}}
\affiliation{Department of Chemistry, University of Florida, Gainesville, FL 32611-7200, USA}
\author{Matthew J. \surname{Andrus}}
\affiliation{Department of Chemistry, University of Florida, Gainesville, FL 32611-7200, USA}
\author{Daniel R. \surname{Talham}}
\email{talham@chem.ufl.edu}
\affiliation{Department of Chemistry, University of Florida, Gainesville, FL 32611-7200, USA}
\author{Mark W. \surname{Meisel}}
\email{meisel@phys.ufl.edu}
\affiliation{Department of Physics and NHMFL, University of Florida, Gainesville, FL 32611-8440, USA}

\begin{abstract}
Cubic heterostructured ({\bf BA}) particles of Prussian blue analogues, composed of a shell of 
ferromagnetic K$_{0.3}$Ni[Cr(CN)$_6$]$_{0.8} \cdot 1.3\;$H$_2$O ({\bf A}), $T_c~\sim~70$~K, 
surrounding a bulk core of photoactive ferrimagnetic Rb$_{0.4}$Co[Fe(CN)$_6$]$_{0.8} \cdot 1.2\;$H$_2$O 
({\bf B}), $T_c~\sim~20$~K, have been studied. Below $T_c~\sim~70$~K, these samples exhibit a persistent 
photoinduced decrease in low-field magnetization, and these results resemble data from other core-shell 
particles and analogous {\bf ABA} heterostructured films. This net decrease suggests that the photoinduced 
lattice expansion in the {\bf B} layer generates a strain-induced decrease in the magnetization of the 
{\bf A} layer, similar to a pressure-induced decrease observed by others in a pure {\bf A}-like material 
and by us in our {\bf BA} cubes. Upon further examination, the data also reveal a significant portion of 
the {\bf A} material whose superexchange, $J$, is perturbed by the photoinduced strain from the {\bf B} 
constituent.
\end{abstract}

\pacs{75.50.Xx, 75.30.Et, 75.70.Cn}

\keywords{Prussian blue analogues, photoinduced magnetism, heterostructures}

\maketitle

\section{INTRODUCTION}
Molecule-based magnets present an intriguing field of materials due to their combination of magnetism 
with various other physical phenomena. Specifically, those materials which exhibit photomagnetic properties 
provide a vast potential for technological applications. In recent years, Prussian blue analogues (PBAs) 
have piqued the interest of the molecular magnetism community due to their range of magnetic properties. 
Although examples exist that are known to order magnetically at or above room temperature 
\cite{Ferlay,Verdaguer}, the feature that distinguishes a particularly appealing subset of PBAs is the 
phenomenon of a charge-transfer-induced spin transition (CTIST) \cite{ShimamotoIC}, which is linked with 
that of persistent photoinduced magnetism (PPIM) \cite{Epstein}. When these properties are combined with 
long-range magnetic order, the potential for information storage applications begins to emerge.

A quintessential example of such a material is the CoFe-PBA, A$_i$Co[Fe(CN)$_6$]$_j \cdot n$H$_2$O, where A 
is a monovalent alkali cation.  CoFe-PBAs are well-known to exhibit a photoinduced CTIST up to a temperature 
of $150$~K \cite{Sato96, Sato99, Bleuzen}, as well as a similar photoinduced transition within the hysteresis 
of the thermal CTIST near, and even at, room temperature for certain stoichiometries 
\cite{Shimamoto, ShimamotoIC}. However, these photoinduced effects are weaker above  $20$~K, which is the 
critical temperature, $T_c$, for the long-range magnetic ordering. The fact that the photoeffect in the 
CoFe-PBA is also structural in nature has been utilized to effect an increase in the temperature at which 
photocontrol is possible, by incorporating this material into heterostructures \cite{Dumont11,Paj10,Paj11} 
with PBAs having higher ordering temperatures and known to be pressure-sensitive \cite{Zentkova07}. To guide 
work pursuing room temperature photocontrolled magnetism, we seek a deeper fundametntal understanding of the 
mechanism leading to these novel effects in the heterostructures, focusing here on quantifying the amount of 
material, in the non-photoactive layers, affected by this photoinduced strain from the CoFe-PBA.

\begin{figure}[b]
\includegraphics[width=0.5\textwidth]{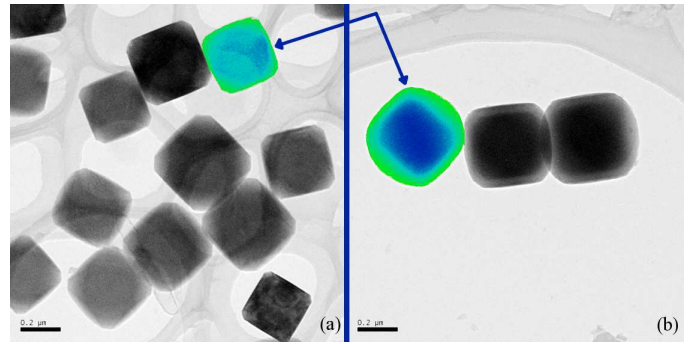}
\caption{(Color online) TEM images of (a) core ({\bf B}) and (b)~core-shell ({\bf BA}). Each scale bar is 
200 nm. The arrows denote a selected particle in each image that has been colorized to highlight the 
core-shell contrast.}
\label{fig1}
\end{figure}

In the present study, core ({\bf B}) and core-shell ({\bf BA}) particles of PBAs (Fig.~\ref{fig1}), composed 
of shells of  K$_{0.3}$Ni[Cr(CN)$_6$]$_{0.8} \cdot 1.3\;$H$_2$O ({\bf A}), which orders ferromagnetically 
with $T_c~\sim~70$~K, surrounding bulk cores of Rb$_{0.4}$Co[Fe(CN)$_6$]$_{0.8} \cdot 1.2\;$H$_2$O ({\bf B}), 
which is photoactive and orders ferrimagnetically with $T_c~\sim~20$~K, have been investigated. 
Below $T_c~\sim~70$~K of the shell, these samples exhibit a persistent photoinduced decrease in low-field 
susceptibility, similar to those results from other core-shell particles \cite{Dumont11} and analogous 
{\bf ABA} heterostructured films \cite{Paj10,Paj11}. This net decrease is attributed to the photoinduced 
lattice expansion in the {\bf B} layer, which generates a strain-induced decrease in the magnetization of 
the {\bf A} layer, similar to a pressure-induced decrease observed by others in a pure {\bf A}-like material, 
attributed to spin-canting \cite{Zentkova07}. Upon further examination, the core-shell susceptibility data 
also reveal a significant portion of the {\bf A} material whose superexchange, $J$, is perturbed with the 
application of this photoinduced strain from the {\bf B} material.

\section{Experimental Details}
\subsection{Synthesis}
K$_3$Cr(CN)$_6$ was synthesized by treating aqueous solutions of potassium cyanide with 
CrCl$_3 \cdot 6\,$H$_2$O and was used after recrystallization from methanol \cite{Bigelow}. Deionized water 
used in synthetic procedures was obtained from a Barnstead NANOpure.  All of the other reagents were 
purchased from Sigma-Aldrich or Fisher-Acros and were used without further purification. The filters used 
during the synthesis were Fast PES Bottle Top Filters with $0.45~\mu$m pore size (Nalgene).

The synthesis of the core particles was performed at room temperature. An aqueous solution ($100$~mL) of 
CoCl$_2 \cdot 6\,$H$_2$O ($0.40$~mmol, $4.0$~mM), and an equal volume of an aqueous solution containing 
K$_3$Fe(CN)$_6$ ($0.45$~mmol, $4.5$~mM) were simultaneously added dropwise to $200$~mL of nanopure water. 
To substitute the Rb cation, RbCl ($0.8$~mmol, $8$~mM) was included in the CoCl$_2$ solution. The final 
solution was kept under vigorous stirring for one hour after all additions were complete. The particles 
were subsequently filtered under vacuum using a $0.45~\mu$m filter before being washed with ultrapure water 
and redispersed in $100$~mL of water for use in the next step.  To isolate the particles, instead of water, 
the particles were concentrated by centrifugation and dried under a flow of air.

To prepare the core-shell particles, the previously synthesized core particles solution was diluted with 
water to 400 mL. An aqueous solution ($50$~mL) of NiCl$_2 \cdot 6\,$H$_2$O ($0.095$~mmol, $1.9$~mM) and an 
aqueous solution ($50$~mL) of K$_3$Cr(CN)$_6$ ($0.105$~mmol, $2.1$~mM) were added dropwise ($10$~mL/hour, 
peristaltic pump) under stirring at room temperature. The final solution was kept under vigorous stirring 
for one hour after all additions were complete. Once the addition was complete, the particles were filtered 
using a $0.45~\mu$m filter and rinsed with ultrapure water.  To isolate the particles, instead of water, the 
particles were concentrated by centrifugation and dried under a flow of air.

\subsection{Instrumentation}
Transmission electron microscopy (TEM) was performed on a JEOL-2010F high-resolution transmission electron 
microscope at $200$~kV. TEM grids used were carbon film on a holey carbon support film, 400 mesh, copper 
from Ted-Pella, Inc. Grids were prepared by dropping $20~\mu$L of a solution containing $5$~mg of sample 
dispersed in $2$~mL of EtOH by sonication for $30$ min. Energy dispersive x-ray spectroscopy (EDS) was 
performed with an Oxford Instruments EDS X-ray Microanalysis System coupled to the TEM.

Room temperature powder x-ray diffraction (XRD) was performed on a Bruker DUO diffractometer. Powder samples 
were packed into $0.3$~mm diameter boron-rich thin walled capillary tubes purchased from the Charles Supper 
Company.  Diffraction patterns were collected in the range $5 ^\circ \leq 2\theta \leq 89 ^\circ$ with a 
$600$~sec/image collection time. The raw data from the CCD detector were converted to powder patterns using 
PowDLL converter.

\begin{figure}
\includegraphics[width=0.5\textwidth]{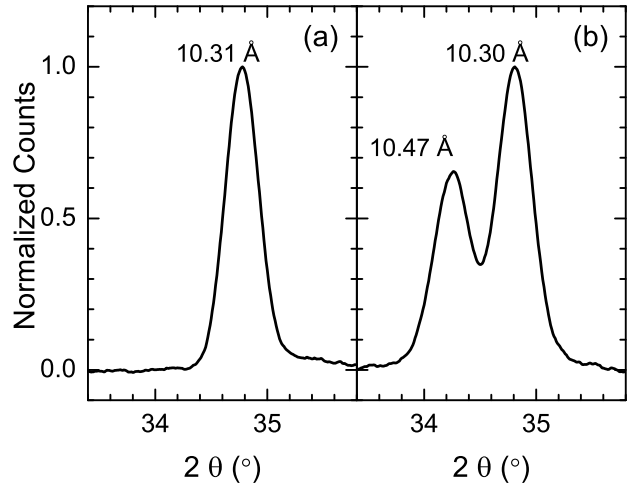}
\caption{The [400] peaks of the XRD spectra for (a) the core ({\bf B}) sample and (b) the core-shell 
({\bf BA}) sample, normalized to the amplitude of the {\bf B} peak in each, as a function of $2 \theta$.}
\label{fig2}
\end{figure}

Fourier-transform infrared (FTIR) spectroscopy was performed on a Nicolet 6700 Thermo Scientific 
spectrophotometer. Powder samples were mixed with KBr and pressed into pellets under $20$~MPa. Scans 
(typically 64) were performed between $2200$ and $1900$~cm$^{-1}$, in the region of the CN stretching bands, 
with a precision of $0.482$~cm$^{-1}$. Corresponding scans of a pure KBr pellet were taken as the background 
measurement.

Magnetization measurements were performed on a Quantum Design MPMS XL-7 superconducting interference device 
(SQUID) magnetometer using a homemade fiber optic sample rod with a bundle of 10 optical fibers to bring 
light into the sample space \cite{Park}. Powder samples were spread between two pieces of transparent tape 
and held in transparent drinking straws directly below the optical fibers. Visible white light was supplied 
by a $150$~W tungsten-halogen lamp located at room temperature. The temperature dependence of the low-field 
susceptibility was measured in an applied field of $100$~G while warming the sample from base temperature, 
$2$~K. Isothermal magnetization was measured at $2$~K while sweeping the field between $70$~kG and $-70$~kG. 
All light state measurements were performed subsequent to 8 hours of isothermal irradiation at $40$~K, after 
irradiation had been ceased. The sample was then cooled again to base temperature, where the same protocol 
that was used for the dark state measurements was repeated to acquire the light state data set.

\begin{figure}
\includegraphics[width=0.5\textwidth]{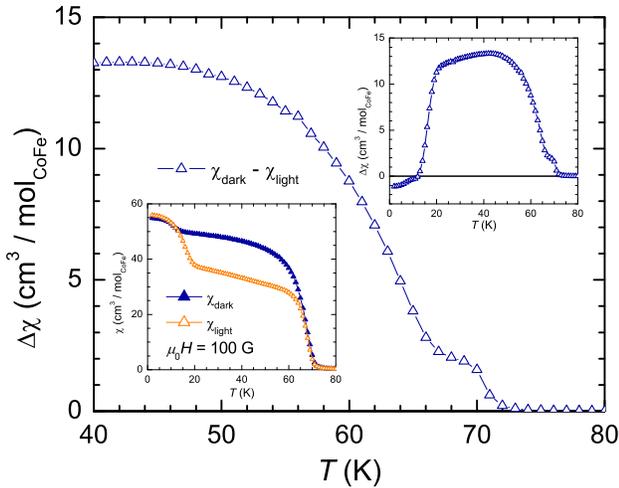}
\caption{(Color online) The difference between susceptibilities in the dark and light states, $\Delta \chi$, 
of the {\bf BA} sample near the transition temperature of the {\bf A} constituent, where the shoulder 
between $66$~K and $72$~K is indicative of the shift in $T_c$ due to the photoinduced strain. The bottom left 
inset shows the untreated susceptibility, $\chi$, in the dark and light states.  The top right inset shows 
$\Delta \chi$ over the entire range of temperatures.}
\label{fig3}
\end{figure}

\section{Results and Discussion}
The TEM images reveal contrast between core and shell in the {\bf BA} particles not present for the pure 
{\bf B} particles (Fig. \ref{fig1}). Additionally, EDS line scans of individual particles show Co and Fe 
content concentrated in the cores, with Ni and Cr content predominately occupying the edges \cite{MFDthesis}. 
The FTIR spectra display a new CN stretching frequency, near $2175$~cm$^{-1}$, arising after addition of the 
{\bf A} shell, consistent with observations for the pure {\bf A} material. Powder XRD patterns also present 
two distinct lattice constants consistent with the single-phase {\bf B} and {\bf A} materials, respectively, 
while only the {\bf B} phase is exhibited in the core particles (Fig. \ref{fig2}). 

The low-field susceptibility of the {\bf BA} sample exhibits a persistent photoinduced decrease 
(Fig. \ref{fig3}). Although the {\bf B} material is a well-established photomagnet \cite{Sato96, Sato99, 
Bleuzen}, these heterostructures invert the sign of the effect in addition to increasing the temperature of 
photocontrol \cite{Dumont11, Paj10, Paj11}.  This photoeffect possesses a significant magnitude up to 
$T_c \sim 70$ K of the {\bf A} material, attributed to a strain-induced canting of the {\bf A} spins, similar 
to pressure-induced effects in a similar material \cite{Zentkova07}. 

\begin{figure}
\includegraphics[width=0.5\textwidth]{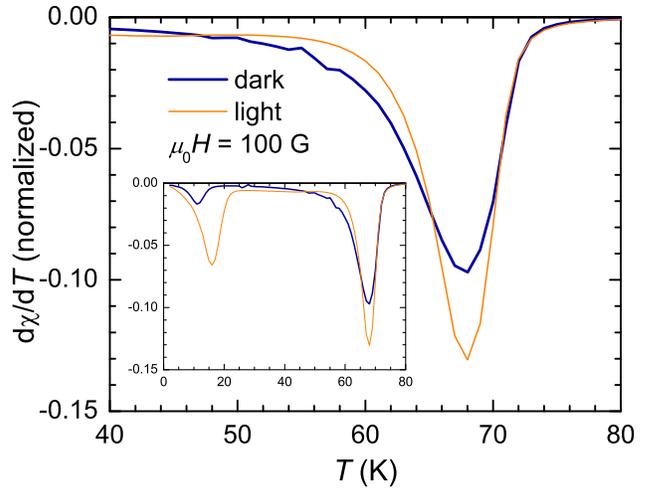}
\caption{(Color online) The derivative of the susceptibilities, $\mathrm{d} \chi / \mathrm{d} T$, in the dark 
and light states near the transition temperature of the \textbf{A} constituent. The inset shows 
$\mathrm{d} \chi / \mathrm{d} T$ over the entire range of temperatures.}
\label{fig4}
\end{figure}

Additionally, a subtle shoulder near the $T_c$ of the {\bf A} material emerges in the photoinduced decrease, 
$\Delta \chi$, indicating a shift in the superexchange of some portion of the shell. In order to quantify the 
amount of affected material, the derivative of the susceptibility was taken and normalized to the same area, 
in the range $40$~K to $80$~K, for each data set (Fig. \ref{fig4}). The validity of normalizing the light and 
dark data sets to equal areas is substantiated by the corresponding saturation magnetizations, which indicate 
no loss of spins upon irradiation. Thus, taking the area under these curves to represent the total number of 
spins in the shell, this area must remain constant for a given sample between the dark and light states. Upon 
inspection of these derivatives, an aspect of the effect not immediately obvious in the raw data 
(Fig. \ref{fig3} inset) becomes apparent. Namely, the magnetic transition of the {\bf A} material sharpens 
significantly when the {\bf B} material is transitioned to the light state. Again, this observation indicates 
that a significant portion of the {\bf A} material experiences a change in its superexchange.

Guided by these observations, we apply a simple model of the shell comprising two distinct regions: an inner, 
strain-sensitive interface region and an outer, rigid bulk-like region, which remains unaffected by switching 
between the dark and light states. Taking these regions to represent distinct superexchange values, the areas 
under the curves in Fig. \ref{fig4}, representing the total number of spins in the shell constituent, must 
account for these two regions. Fitting these curves with two peaks, corresponding to the two assumed 
superexchange values, provides a measure of the percentage of the shell in each region. The bulk-like value 
is taken to be the higher $T_c$, which does not shift, and this value remains constant between the fits, 
while the total area is constrained to unity. These fits reveal 50\% of the material with each $T_c$, where 
the bulk-like material has $T_c^{bulk} \simeq 68$~K and the strained material has $T_c^{dark} \simeq 62$~K 
shifting to $T_c^{light} \simeq 65$~K, which is closer to that attributed to the bulk. 

The direction of this shift may be explained phenomenologically in the following manner. Upon cooling, the 
{\bf B} material undergoes a lattice contraction from a $10.3$~\AA \ lattice constant to $10.0$~\AA, thus 
stretching the {\bf A} lattice near the interface. With irradiation, the {\bf B} lattice again expands to 
its room temperature size of $10.3$~\AA, allowing the interfacial {\bf A} lattice to relax closer to its bulk 
state.

\section{CONCLUSION}
By analyzing the first derivative of the susceptibilities in the dark and light states of the {\bf BA} 
particles, we have quantified the amount of the {\bf A} material affected by the strain arising from the 
photinduced expansion of the {\bf B} lattice. Although the shift in superexchange is found to be small 
($\Delta T_c \simeq 3$ K), nearly {\it half} of the volume of the {\bf A} shell experiences this 
superexchange perturbation upon photoswitching the {\bf B} core. A forthcoming report will detail the 
investigation of the same cores with increasing shell thicknesses.

\begin{acknowledgments}
This work was supported by NSF DMR-1005581 (DRT), DMR-1202033 (MWM), the National High Magnetic Field 
Laboratory via cooperative agreement NSF DMR-0654118, and the state of Florida. The authors would like to 
thank Dr. Khalil A. Abboud, for the use of the Bruker DUO instrument, and Kerry Siebein, at the University 
of Florida Major Analytical Instrumentation Center, for TEM and EDS work. 
\end{acknowledgments}



\begin{references}

\bibitem{Ferlay}S.~Ferlay, T.~Mallah, R.~Ouah\`es, P.~Veillet, M.~Verdaguer, 
Inorg. Chem. {\bf 38}, 229-234 (1999).
\bibitem{Verdaguer}M.~Verdaguer, A.~Bleuzen, C.~Train, R.~Garde, F.~F.~Biani, 
C.~Desplanches, Phil. Trans. R. Soc. A {\bf 357}, 2959-2976 (1999).
\bibitem{ShimamotoIC}N.~Shimamoto, S.~Ohkoshi, O.~Sato, K.~Hashimoto, 
\mbox{Inorg.} Chem. {\bf 41}, 678-684 (2002).
\bibitem{Epstein}D.~A.~Pejacovi\'c, J.~L.~Manson, J.~S.~Miller, A.~J.~Epstein, 
Phys. Rev. Lett. {\bf 85}, 1994-1997 (2000).
\bibitem{Sato96} O.~Sato, T.~Iyoda, A.~Fujishima, K.~Hashimoto,  Science 
{\bf 272}, 704-705 (1996).
\bibitem{Sato99}O.~Sato, Y.~Einaga, A.~Fujishima, K.~Hashimoto,  Inorg. Chem. 
{\bf 38}, 4405-4412 (1999).
\bibitem{Bleuzen}A.~Bleuzen, C.~Lomenech,V.~Escax, F.~Villain, F.~Varret, 
C.~Cartier dit Moulin, M.~Verdaguer, J. Am. Chem. Soc. {\bf 122}, 6648-6652 
(2000).
\bibitem{Shimamoto}N.~Shimamoto, S.~Ohkoshi, O.~Sato, K.~Hashimoto, Chem. 
Lett. {\bf 4}, 486-487 (2002).
\bibitem{Paj10} D.~M.~Pajerowski, M.~J.~Andrus, J.~E.~Gardner, E.~S.~Knowles, 
M.~W.~Meisel, D.~R.~Talham,  J. Am. Chem. Soc. {\bf 132}, 4058-4059 (2010).
\bibitem{Dumont11} M.~F.~Dumont, E.~S.~Knowles, A.~Guiet, D.~M.~Pajerowski, 
A.~Gomez, S.~W.~Kycia, M.~W.~Meisel, D.~R.~Talham,  Inorg. Chem. {\bf 50}, 
4295-4300 (2011).
\bibitem{Paj11} D.~M.~Pajerowski, J.~E.~Gardner, F.~A.~Frye, M.~J.~Andrus, 
M.~F.~Dumont,  E.~S.~Knowles, M.~W.~Meisel,  D.~R.~Talham,  Chem. Mater. 
{\bf 23}, 3045-3053 (2011).
\bibitem{Zentkova07} M.~Zentkov\'a, Z.~Arnold, J.~Kamar\'ad, 
V.~Kave\v{c}ansk\'y, M.~Luk\'a\v{c}ov\'a, S.~Mat$'$a\v{s}, M.~Mihalik, 
Z.Mitr\'oov\'a, A.~Zentko,  J. Phys.: Condens. Matter {\bf 19}, 266217 (2007).
\bibitem{Bigelow} J.~H.~Bigelow, Inorg. Synth. {\bf 2}, 203-205 (1946).
\bibitem{Park} J.-H.~Park, Ph.D. Dissertation, University of Florida (2006), 
\url{http://etd.fcla.edu/UF/UFE0013792/park_j.pdf}.
\bibitem{MFDthesis} M.~F.~Dumont, Ph.D. Dissertation, University of Florida 
(2011), \url{http://etd.fcla.edu/UF/UFE0042791/dumont_m.pdf}.
\end{references}
\end{document}